\documentclass[]{spie}  

\usepackage{amsmath,amsfonts,amssymb}
\usepackage{siunitx}
\usepackage{graphicx}

\usepackage{caption}
\usepackage{subcaption}
\usepackage{gensymb}
\usepackage{float}

\usepackage{lineno}

\usepackage[colorlinks=true, allcolors=blue]{hyperref}

\title{Facility integration of the NASA IRTF adaptive secondary mirror}

\author[a]{Ellen Lee}
\author[a]{Mark Chun}
\author[a]{Michael Connelley}
\author[a]{Ruihan Zhang}
\author[b]{Olivier Lai}
\author[a]{Tony Denault}
\author[a]{John Rayner}
\author[c]{Max Baeten}
\author[c]{Arjo Bos}
\author[c]{Matias Kidron}
\author[c]{Fred Kamphues}
\author[c]{Stefan Kuiper}
\author[c]{Wouter Jonker}
\author[a]{Alan Ryan}
\author[d]{Philip Hinz}
\affil[a]{University of Hawai'i at Manoa - Institute for Astronomy, Honolulu, HI, USA}
\affil[b]{Observatoire de la Côte d'Azur, Nice, France}
\affil[c]{TNO, Delft, Netherlands}
\affil[d]{University of California - Santa Cruz, Santa Cruz, CA, USA}

\authorinfo{Further author information: (Send correspondence to E.L.)\\E.L.: E-mail: ellenlee@hawaii.edu\\M.C.: E-mail: markchun@hawaii.edu}

\begin{document} 
\maketitle

\begin{abstract}
IRTF-ASM-1 has been functioning well since its first light in 2024. This adaptive secondary mirror (ASM) was primarily developed to be an on-sky demonstration of the new hybrid variable reluctance actuator technology at the NASA Infrared Telescope Facility (IRTF). However, due to its physical robustness and our previous demonstrations of sensitivity enhancements with the ASM, we are interested in using it to optimize telescope image quality on a nightly basis. This will directly benefit science observations. However, as IRTF does not currently have adaptive optics expertise, we have been developing the system to be used with minimal human intervention. We present our progress in developing software for active optics mode with IRTF-ASM-1 using the single conjugate, facility $2\times2$ Shack-Hartmann wavefront sensor Felix. We also present techniques for removing low-order, large amplitude non-common path aberrations between Felix and our science instruments.
\end{abstract}


\section{Introduction}
IRTF-ASM-1 is a prototype adaptive secondary mirror (ASM) developed for the NASA Infrared Telescope Facility (IRTF), a 3.2-meter infrared-optimized telescope on Maunakea. IRTF-ASM-1 was primarily intended as an on-sky demonstration of the hybrid variable reluctance (HVR) actuator technology developed by the Netherlands Organization for Applied Scientific Research (TNO). The drastically increased efficiency and high linear range of force output of HVR actuators reduce the design complexity of ASMs. This permits a more physically robust design for large ASMs. The HVR technology may also make ASMs more accessible to mid-sized observatories if a reduced number of actuators are used, due to both the robustness and lower cost of acquisition and maintenance.

We successfully had first light with the ASM in April 2024, just 16 months after funding was secured for the project.\cite{lee2024first} We have since brought it to the telescope for approximately one week per semester. At the time of writing, it is not permanently installed because the ASM is taller than IRTF's static secondary mirrors. It does not fit on the hexapod that we use to focus and collimate the telescope. We are working to acquire a new hexapod that will accommodate the ASM within about a year, after which point it may be left on the telescope permanently. The timeline for the hexapod also depends on accommodating a potential second IRTF ASM that will test the next-generation HVR actuators, silicon carbide body, and integrated electronics developed by UC Santa Cruz for the planned Keck I adaptive secondary mirror.\cite{hinz2024keck,kuiper2023developments}

Due to its physical robustness and reliable performance, the goal of the IRTF-ASM-1 project has evolved since first light. It has served as a pathfinder for demonstrating calibrations for larger ASMs, such as on-sky generation of interaction matrices through atmospheric turbulence.\cite{zhang2026cacofoni} In Fall 2025, we successfully integrated the ASM with the facility acquisition camera and 2$\times2$ Shack-Hartmann wavefront sensor Felix. We demonstrated that an off-axis, single-conjugate enhanced seeing mode with Felix is able to improve the throughput of one of our facility slit spectrographs by a factor of 1.6 when using a 0."3 slit.\cite{lee2025ao4elt} Since then, our work has been focused on developing the software infrastructure necessary to operate the ASM without input from an adaptive optics (AO) expert. We report our progress in the facility integration of IRTF-ASM-1 along with on-sky results from a week of engineering time in late May 2026.

\section{System overview}
The layout of the system is illustrated and described in detail in Ref.~\citenum{lee2025ao4elt}. To summarize: IRTF-ASM-1, its drive electronics, and the real-time controller (RTC) for our test $12\times12$ wavefront sensor (WFS) are mounted on the top end ring. This test WFS is Cassegrain-mounted, as are all of IRTF's science instruments, meaning that it must be removed from the center position if we want to use the ASM with a science instrument. As a result, we use Felix to control the ASM when performing science observations. Felix is mounted above the Cassegrain port and uses a pickoff mirror to acquire off-axis stars between \qty{2.5}{\arcminute} and \qty{4}{\arcminute} away from the science target. The active optics system at IRTF uses Felix to measure low-order Zernike modes through coma.

We will now discuss our envisioned usage of IRTF-ASM-1 and software developments for integration with the observatory. This project has been opportunistic, from our initial decision to go on-sky at IRTF due to the small size of its secondary mirror to its integration with Felix, which was serendipitously installed within months of first light with the ASM. IRTF-ASM-1 has since been used to develop calibration techniques for larger ASMs; since Fall 2025, it has started to evolve into a pathfinder for retrofitting enhanced seeing capability on older telescopes to improve the throughput of facility instruments. The decision to permanently install the ASM was a relatively recent development, occurring somewhat nebulously after we had gained confidence that it was robust through our past few semesters of observing. For these reasons, the present state of the system is not the result of strict performance requirements from IRTF. The design of the software is motivated not only by its long-term usage for science, but also by ease of access by other researchers and students to develop on-sky calibration techniques in the future.

In the near term, our goal is to use the ASM to remove quasi-static aberrations that mainly arise from the primary mirror support structure of IRTF. These aberrations are known to degrade the PSF by over a factor of two in FWHM when the telescope is pointed two airmasses in any of the cardinal directions\cite{dinh2024measuring}, making them a potential area of major improvement. Nevertheless, due to the aforementioned reasons, the software is designed to accommodate an enhanced seeing mode, which shares implementation with active optics. It may be activated manually by increasing the frame rate of the Felix camera, which sets the loop rate. The high-level design of the software is illustrated in Fig.~\ref{fig:software}. (Despite not being true adaptive optics, the active optics software is labeled "AO software" because it also implements enhanced seeing mode, aside from the automation of setting the exposure time and loop parameters.) The function of each component is as follows:
\begin{figure}[htpb]
    \centering
    \includegraphics[width=0.75\linewidth]{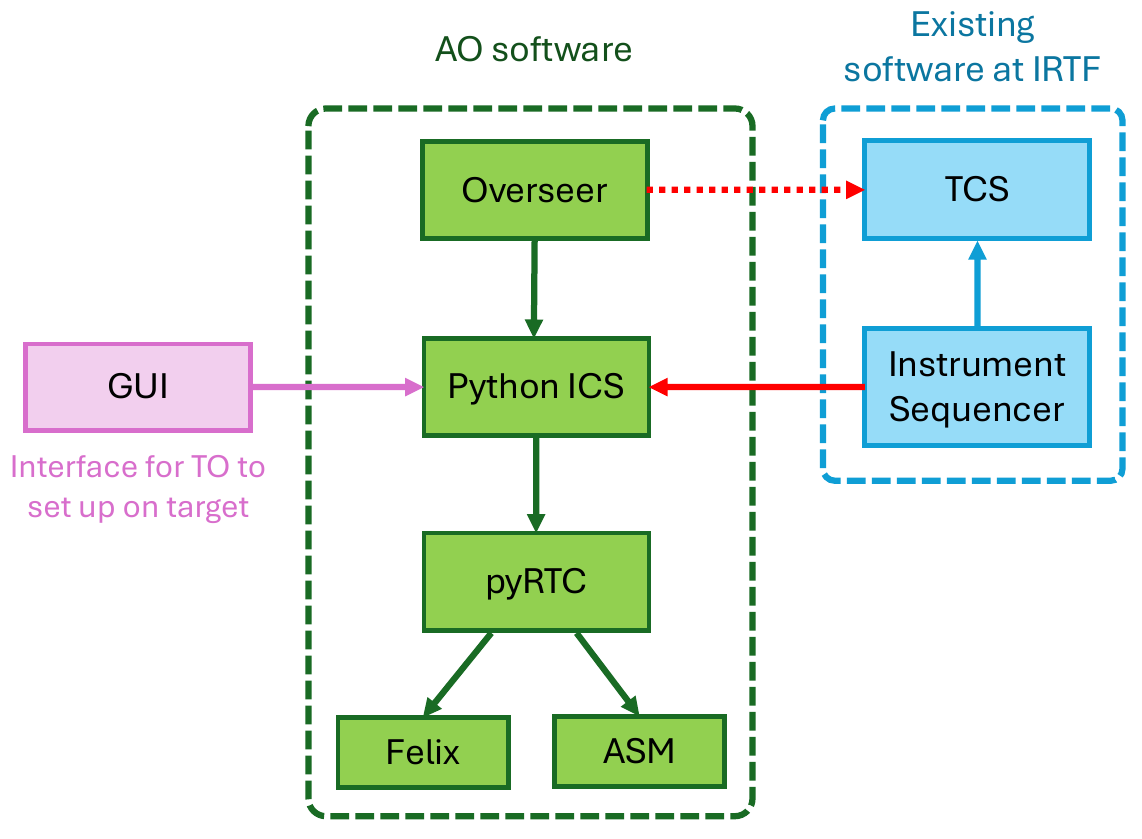}
    \caption{Design of the software for active optics with IRTF-ASM-1, which can also be used to operate an enhanced seeing mode. The telescope operator (TO) interacts with the system via a GUI used for target acquisition. See the text for details.}
    \label{fig:software}
\end{figure}
\begin{itemize}
    \item \textbf{Hardware - Felix camera and the ASM:} Commands are transmitted to the ASM via local internet. We use the official Python wrapper for the Andor Software Development Kit to communicate with the Felix camera. A minor annoyance is that only a single application can access the camera at a time. This prevents us from accessing active optics at the same time as the existing Felix camera control software that is currently used to point the telescope, which is written in C.
    \item \textbf{pyRTC:} As described in Ref.~\citenum{lee2025ao4elt}, our motivation behind using pyRTC\footnote{\url{https://github.com/jacotay7/pyRTC}} as the real-time controller is its user-friendliness and our permissive requirements for the loop rate. pyRTC includes routines for measuring interaction matrices (e.g., DO-CRIME), loading slope offsets, and other calibrations.
    \item \textbf{Python ICS:} The Python Instrument Control Server (ICS) for active optics mode receives commands from other modules and appropriately relays them to pyRTC. This includes abstraction for opening and closing the loop, starting and shutting down hardware, etc.
    \item \textbf{GUI:} This interface is the point of interaction for the telescope operator (TO). The GUI is used to acquire the target by setting the location of the subaperture masks for each of the four spots in pyRTC. It can also be used to manually change the loop gains and exposure time, although it should not be necessary to touch these parameters.
    \item \textbf{Instrument sequencer:} For each instrument, a sequencer sends commands to the telescope (and/or other components) based on observing parameters set by the user; for example, to command a nod of the telescope if the user desires an AB pair. In this case, the sequencer will also ask the ICS to open the loop, update the subaperture mask positions, and close the loop again.
    \item \textbf{TCS:} We communicate with the telescope control system (TCS) to offload tip-tilt to telescope pointing and to acquire the AB positions before a beamswitch occurs.
    \item \textbf{Overseer:} The overseer checks the system at some pre-determined frequency--\qty{10}{\hertz} during our on-sky tests--and sends commands to pyRTC via the Python ICS if certain criteria are met. It is currently used for tip-tilt offloading, but may also be used to perform safety checks such as monitoring the signal-to-noise ratio of the spots.
\end{itemize}

We note that our initial implementation was more similar to what IRTF currently implements for its guiders and acquisition cameras, in which they query the TCS prior to every exposure rather than receiving knowledge of the state of the system from the instrument sequencer. An advantage of this is that it fully decouples active optics from IRTF's instrument software and thus does not require modification of the sequencer. However, this approach is not extensible for a long-term enhanced seeing mode because the TCS was not designed to be queried robustly at high temporal frequencies.

\section{Calibrations}
In practice, running the system amounts to finding a guide star within the patrol field of Felix and using the GUI to locate the position of the spots on the detector. We plan to set the gain and exposure time of Felix automatically depending on the brightness of the off-axis star, with other commands during the observation being handled by the sequencer. This section describes our processes for acquiring necessary calibrations to close the loop.

\subsection{Field curvature}
Non-common path aberrations (NCPAs) between Felix and the instrument plane are our largest inhibitor. NCPAs limited our ability to test the improvement in image quality in enhanced seeing mode, as our measured improvement in throughput over the natural seeing was strongly contingent upon removing the NCPAs manually. Focus comprises the largest error term in these NCPAs.

There is field curvature across the patrol field of Felix because IRTF is a classical Cassegrain telescope. To calibrate the field curvature, we manually translated the ASM along the optical axis using the focusing mechanism, which allowed us to measure a scaling factor between the physical displacement and the focus coefficient measured in Felix. The focus coefficient was extracted using the slopes measured with Felix and a modal interaction matrix. Based on the optical prescription of IRTF, the field curvature was known to amount to \qty{23}{\milli\meter} of displacement towards the secondary mirror between the center of the field and an outer radius of \qty{270}{\arcsecond}, which thereby yielded a model of the focus coefficient offset versus off-axis guide star position.

\begin{figure}[htpb]
    \centering
    \includegraphics[width=\linewidth]{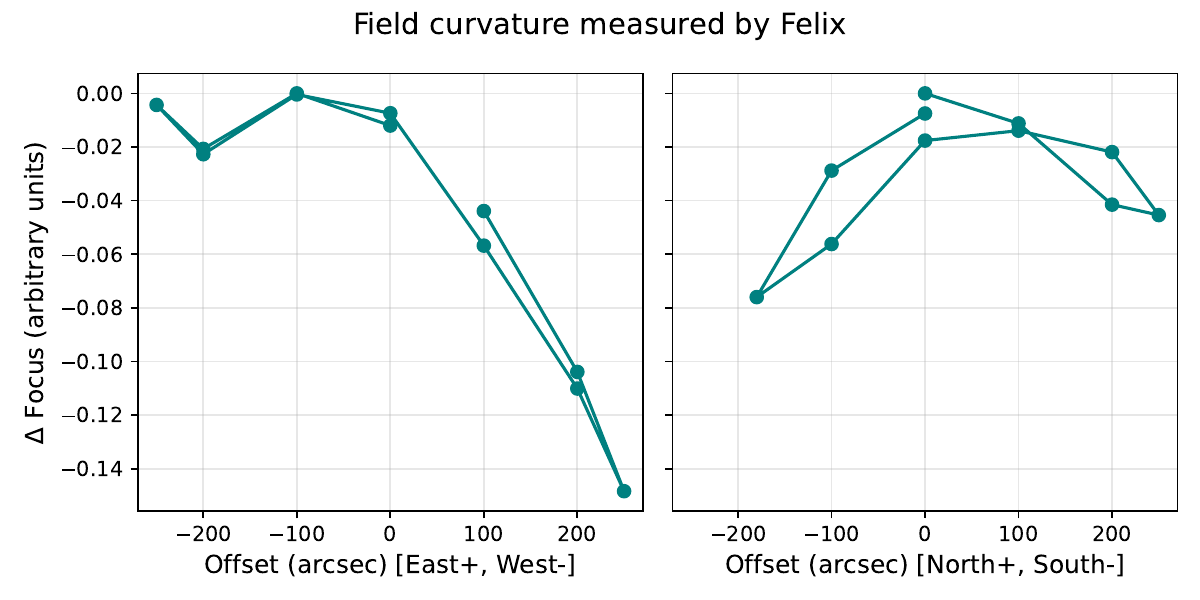}
    \caption{Focus values measured in Felix for different off-axis wavefront sensor star positions. The field curvature is expected to follow a parabola. \textit{Left:} It is apparent that the focal plane is decentered by about \qty{150}{\arcsecond} to the west. \textit{Right:} No such offset is seen in the north-south direction.}
    \label{fig:apos}
\end{figure}

We also obtained an empirical measurement of the field curvature without the ASM on the telescope. Starting with a measurement of the focus coefficient in Felix with the star on-axis, we moved the telescope in each of the cardinal directions and measured the focus coefficient for each offset value. Figure~\ref{fig:apos} shows that field curvature is indeed present. We can also see that the focal plane of IRTF is offset laterally relative to the reference axis, which is defined by the instrument rotator bearing. This is likely due to our procedure for collimating IRTF. The final step requires us to tilt the secondary mirror to remove the on-axis coma, steering the optical axis off of the reference.

We were limited in our ability to test the theoretical lookup table due to poor seeing during our engineering run in 2026A.\footnote{Most of the seeing was at higher altitudes, so we could not enhance the seeing with Felix.} We plan to test the quality of the lookup table further in the future.

\subsection{NCPA measurement in frequency space} \label{sec:calib.ncpa}
Motivated by the calibration challenges with enhanced seeing mode, we have developed an empirical focal plane sharpening technique that works in the regime of large aberrations. Details regarding how the method functions and the choice of a figure of merit will be described in a forthcoming paper, but a brief explanation will be provided here. This approach is likely not necessary with active optics mode at IRTF unless the seeing is exceptional.

\begin{figure}[htpb]
    \centering
    \includegraphics[width=0.85\linewidth]{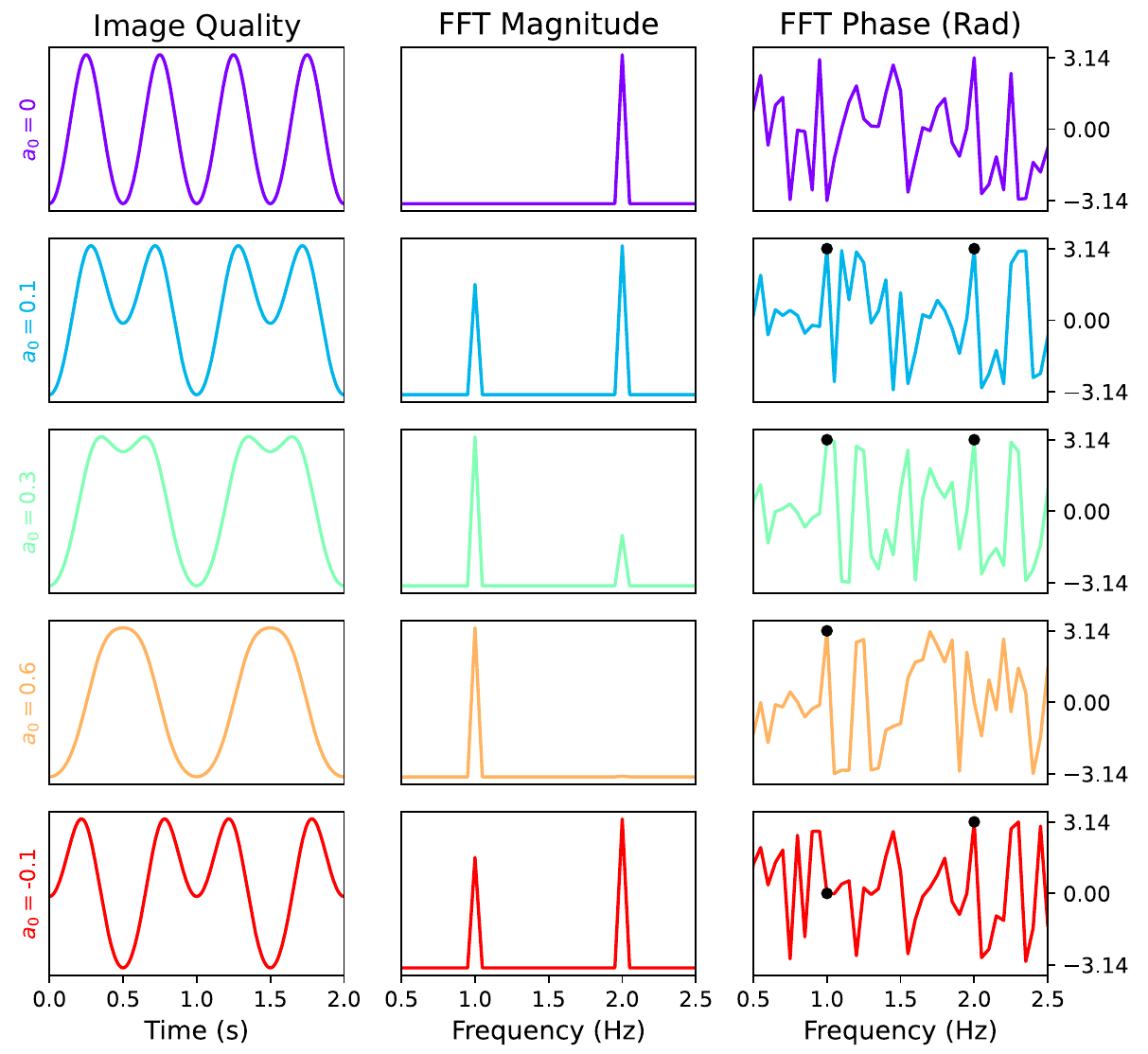}
    \caption{Simulated NCPA offset measurement for a single mode modulated at \qty{1}{\hertz} with an amplitude of 0.5. Each row represents a different starting NCPA offset ($a_0$ labeled on the left; also indicated by unique colors). The left column shows the measured image quality, the middle shows the magnitudes of the fast Fourier transform of the image quality, and the right shows the phases. The image quality oscillates at twice the injected frequency when the starting offset is zero (top row, purple) and at one times the frequency when the offset is outside the capture range (fourth row, yellow). Between these extremes, we see a combination of both frequency components. The sign of the coefficient can be extracted from the phases of the components (right column) as long as the time delay between the injected signal and measured image quality is known.}
    \label{fig:ncpa.examples}
\end{figure}

The method works by modulating orthogonal spatial modes at unique input frequencies, recording fast focal plane images, and then extracting the response of the image to each mode in Fourier space. The principle is illustrated in Fig.~\ref{fig:ncpa.examples}. For example, imagine modulating a single Zernike mode at 1 Hz, as a sine wave with an amplitude of one and centered about zero. Then, quickly record focal plane images and measure the image quality for each image. If the correct NCPA offset to optimize the image quality is, say, +1.5, then the image quality will oscillate at \qty{1}{\hertz}, with the peak being whenever the input is -1. However, if the ideal offset is zero, then the image quality will oscillate at \qty{2}{\hertz} because it can only worsen, regardless of the sign of the input. This insight is based on a method developed by Ref.~\citenum{woillez2019naomi} and is shown in the first and fourth rows of Fig.~\ref{fig:ncpa.examples}.

Now, imagine a case where an NCPA offset is present, but not so large that it is outside the capture range of modulation. The measured image quality is a combination of both the 1 and \qty{2}{\hertz} components, with more of the \qty{2}{\hertz} component the closer the image quality is to being optimal (see rows two, three, and five in Fig.~\ref{fig:ncpa.examples}). This is because the measured image quality is necessarily periodic within the injected modulation frequency of \qty{1}{\hertz} and is therefore described by a Fourier series, i.e., a linear combination of harmonics of the injected frequency. So by viewing the measured image quality in frequency space, we can obtain a very good approximation of the NCPA offset from the ratio of the magnitudes of the 1 and \qty{2}{\hertz} components (and additional harmonics if desired, although there is relatively little power here). The sign of the coefficient comes from the phase of each frequency component.

\begin{figure}[htpb]
    \centering
    \includegraphics[width=0.6\linewidth]{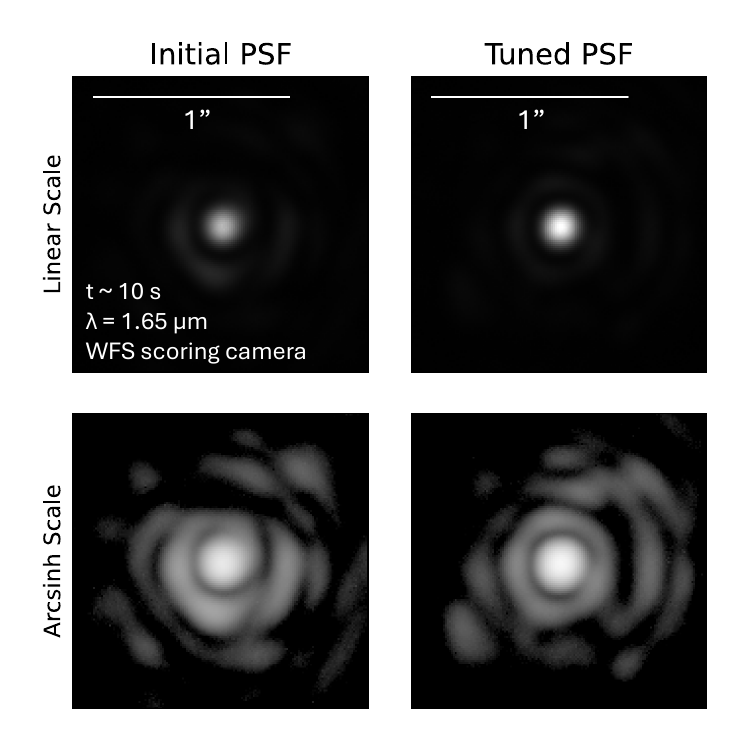}
    \caption{On-sky improvement in image quality using the NCPA tuning technique described in Sec.~\ref{sec:calib.ncpa}, tuning individual modes with a single iteration each. Images are of Hokule'a (alf Boo) using the infrared scoring camera on the $12\times12$ wavefront sensor breadboard. The Strehl ratio was increased from 27\% to 37\%, with all of the correction being in focus and coma. We did not find a noticeable astigmatism offset.}
    \label{fig:ncpa_sky}
\end{figure}

\begin{figure}[htpb]
    \centering
    \includegraphics[width=0.85\linewidth]{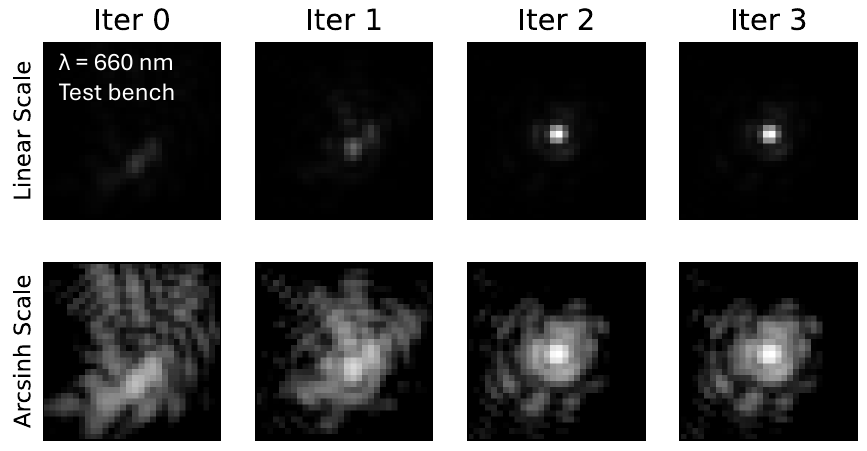}
    \caption{Lab improvement in image quality, tuning both astigmatisms and comas at once using \qty{20}{\second} of data per iteration. All of the improvement was seen within the first two iterations, with the second iteration only being necessary due to the initial NCPA offsets purposefully being set to lie outside of the capture range of modulation. We aim to demonstrate this dynamic range on-sky in open loop to help flatten large ASMs during first light.}
    \label{fig:ncpa_lab}
\end{figure}

We verified the basic working principle of this method on-sky by sequentially measuring the focus, astigmatism, and coma offsets using a \qty{1}{\hertz} probe (Fig.~\ref{fig:ncpa_sky}). The NCPA offset for each mode was extracted from 5 to \qty{10}{\second} of data each, without iteration. Our plan is to use this method at the focal plane of our science instruments to improve image quality in enhanced seeing mode with Felix. In the future, we would like to verify certain advantages of the method on-sky, including its high dynamic range and modal multiplexing (validated in the lab, shown in Fig.~\ref{fig:ncpa_lab}) along with its potential to work in open loop. The latter will be useful for flattening the University of Hawai'i 88-inch telescope's ASM, for which we do not have a test bench to develop a set of flat commands in the lab before bringing it on-sky.

\subsection{Nightly calibration procedure}
Of the calibrations required to close the loop, there are three that may be variable: 1) the interaction matrix, 2) the high-order flat shape of the ASM, and 3) the NCPA offset. Due to Felix having two subapertures across the width of the pupil, the interaction matrix is forgiving unless the rotation of the ASM has changed. We were able to obtain stable closed loop performance with an empirical DO-CRIME\cite{lai2021crime} interaction matrix that we measured in Oct 2025. Thus, we usually will not need to update the interaction matrix. The high-order flat shape of the ASM similarly appears to be very stable across two years and many telescope pointings\cite{lee2025ao4elt}. We therefore do not consider it to be a necessary nightly calibration, but we would still like to monitor the high-order shape over long periods of time. We will not discuss methods for doing this in depth, but this may involve removing the ASM from the telescope and bringing it to the lab setup. It may also be possible to implement curvature wavefront sensing by chopping the ASM and applying different amounts of focus to each end of the chop.

The NCPA offset consists of two components. First, each instrument has a different focus position with respect to the telescope, which should hypothetically not change from night to night. In practice, the best focus value for each instrument appears to drift from night to night. We are investigating why this is the case. In the meantime, we may begin with the following calibration at the beginning of the night and whenever instrument changes are performed:
\begin{enumerate}
    \item Point to a star near zenith and acquire an off-axis guide star with Felix.
    \item Manually focus the image in the slit viewer of the science instrument by translating the secondary mirror. The TO may follow an existing procedure to do this.
    \item Measure the slopes in Felix in open loop, averaging over \qty{60}{\second} of data.
    \item Use the interaction matrix to convert these slopes to Zernike coefficients and extract the focus term.
    \item Update the baseline reference slopes in Felix to use the measured focus coefficient.
\end{enumerate}

The second part of the NCPA offset is a variable component that is likely the result of flexure, and results in terms that are higher order than focus (e.g., astigmatism and coma). Although these terms are necessary to calibrate to maximize the performance of enhanced seeing mode, we have not yet characterized how quickly they evolve and whether they can be compensated via a lookup table.

If an enhanced seeing mode is integrated in the future, some form of gain optimization may be necessary. Because the spots in Felix are quite oversampled, we are not concerned with the optical gain. This means that we may use a model that depends on guide star brightness, loop rate, seeing conditions, etc. It may be desirable to implement a separate gain for the tip/tilt for fast guiding without correcting higher order terms. It would also be helpful if for some reason we need to guide the telescope normally while still compensating for aberrations in the primary mirror. 

\section{Status and schedule}

\begin{figure}[htpb]
    \centering
    \includegraphics[width=0.65\linewidth]{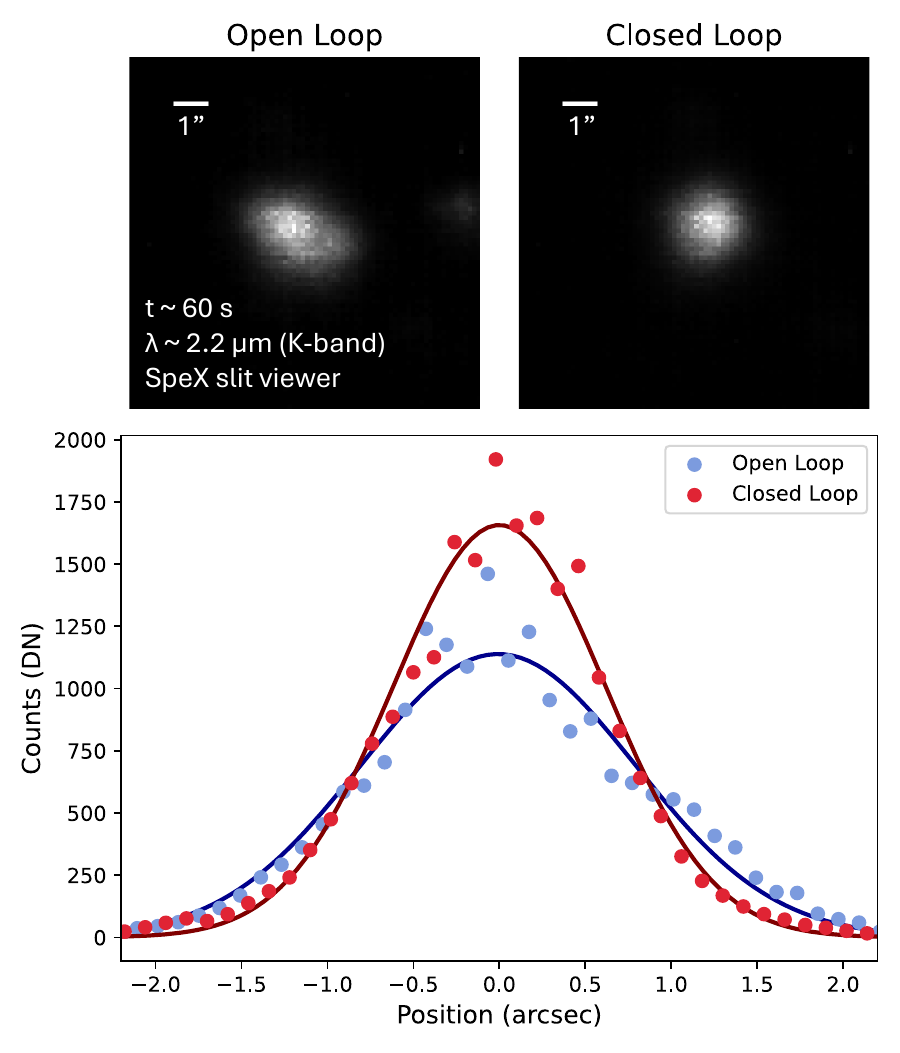}
    \caption{A basic demonstration of active optics mode in SpeX. The target is 2.5 airmasses to the south (Shaula, or lam Sco). Despite the poor seeing, there is clearly severe astigmatism present in the primary mirror. \textit{Top:} Slit viewer images in SpeX in K-band (2.027 to \qty{2.363}{\micro\meter}). There is a factor of 1.4 improvement in the equivalent-noise area. The faint smudge to the right of the left panel is a field star. \textit{Bottom:} A cut-through of the images on the top along the axis of the slit.}
    \label{fig:active_guidedog}
\end{figure}

Figure~\ref{fig:active_guidedog} shows a demonstration of active optics with SpeX, a low- to mid-resolution near-infrared spectrograph and the most popular instrument at IRTF.\cite{rayner2003spex} The images in the slit viewer show an improvement by a factor of 1.4 in equivalent-noise area\cite{1983PASP...95..163K} at high airmass, in mediocre seeing conditions. This is separate from the improvement in sensitivity that we measured from enhanced seeing mode, for which we removed static aberrations in the telescope before recording open loop data.

The current implementation can cover nearly all observing modes at IRTF, which generally use some combination of sidereal/non-sidereal tracking and A/AB nodding. The ASM can be used as a static secondary mirror if a guide star is not available. There is currently no integration for mid-infrared chopping, which requires very large amplitudes of tilt (several arcseconds for point sources) at \qty{4}{\hertz}. We have previously demonstrated that the ASM is fundamentally capable of chopping on-sky.\cite{lee2025ao4elt} A notable edge case that cannot be covered by the ASM is when the chop amplitude is many arcseconds, which may be necessary for mid-infrared planetary observations. This requires a separate chop mechanism and replacement of the ASM with a static secondary mirror, as we do not want to rapidly shake the ASM.

We have additional engineering time in late October 2026. Although we have demonstrated the software at a basic level and have manually tested calibrations, our goal is to improve the interface to the point where IRTF telescope operators can run the system. Rather than being a contiguous week of nights, as has been typical for the past few semesters, the engineering time is interspersed with several nights of science observations without removing the ASM from the telescope. The ASM will be used with a static shape during these observations. Our goal is to have active optics fully functional by late 2026B, with it being used for regular science observations in 2027.

\section{Conclusions and future work}
We have developed the software architecture for using IRTF-ASM-1 for active optics. We have also demonstrated calibration techniques that will be used to close the loop with the ASM without the $12\times12$ wavefront sensor. The system is currently in a state where it can be run by a non-AO expert, but requires further development until it can be used almost autonomously. Most of this software development involves completing the user interface, automating the exposure time of Felix based on guide star magnitude, and adding safety measures. Thus far, this project has primarily been supported by a graduate student. We are developing documentation and aim to gracefully hand off the system over the next two semesters so that it can be cared for long term by IRTF staff.

As mentioned in the introduction, we have previously demonstrated that an enhanced seeing mode with Felix can also further improve the sensitivity of IRTF. However, there are currently no concrete plans to finish developing enhanced seeing mode, primarily due to the stricter requirements in verifying that atmospheric conditions are suitable and in safely automating the operation of the loop when the system is running at higher temporal frequencies. Such complications may be remedied, but this mode of operation would require IRTF to hire staff with AO expertise to properly develop and care for the system. IRTF is currently seeking feedback from its community to develop a scientific strategic plan, which may determine whether additional resources will be allocated to supporting new capabilities with IRTF-ASM-1.

\acknowledgments 
This work and two of its authors (Lee and Zhang) have been supported by a National Science Foundation Advanced Technology and Instrumentation award (NSF-1910552). The Infrared Telescope Facility is operated by the University of Hawaii under contract 80HQTR19D0030 with the National Aeronautics and Space Administration. The first author of this manuscript is currently funded by the NASA Infrared Telescope Facility. We also wish to recognize and acknowledge the very significant cultural role and reverence that the summit of Maunakea has always had within the indigenous Hawaiian community. The opportunity to conduct observations from this mountain is an enormous privilege, and we feel gratitude for our ability to study astronomy from Maunakea. 

\bibliography{report} 

\begin{thebibliography}{10}

\bibitem{lee2024first}
Lee, E., Chun, M., Lai, O., Zhang, R., Baeten, M., Bos, A., Kidron, M., Kamphues, F., Kuiper, S., Jonker, W., et~al., ``First laboratory and on-sky results of an adaptive secondary mirror with tno-style actuators on the nasa infrared telescope facility,'' in [{\em Adaptive Optics Systems IX}{\nolinebreak\hspace{0.1em}]},   {\bf 13097},  652--666, SPIE (2024).

\bibitem{hinz2024keck}
Hinz, P.~M., Holden, B., Stelter, R.~D., Radovan, M., Hunter, A., Savage, M., Kupke, R., Dillon, D., Lu, J., Chun, M., et~al., ``Keck adaptive secondary mirror overview,'' in [{\em Adaptive Optics Systems IX}{\nolinebreak\hspace{0.1em}]},   {\bf 13097},  639--651, SPIE (2024).

\bibitem{kuiper2023developments}
Kuiper, S., Baeten, M., Maniscalco, M., Bos, A., Jonker, W., Kamphues, F., Chun, M., Hinz, P., and Cruz, S., ``Developments towards an adaptive secondary mirror for keck,'' in [{\em Adaptive Optics for Extremely Large Telescopes 7th Edition}{\nolinebreak\hspace{0.1em}]},  (2023).

\bibitem{zhang2026cacofoni}
Zhang, R., Lai, O., Peck, B., Chun, M., and Lee, E., ``Cacofoni: Cosine amplitude calibration organized in frequency for on-sky nimble interaction-matrix. a novel method for calibrating adaptive optics systems ii.,'' {\em The Astronomical Journal}  (2026).
\newblock In review.

\bibitem{lee2025ao4elt}
Lee, E., Chun, M., Zhang, R., Lai, O., Connelley, M., Denault, T., Taylor, J., Rayner, J., Baeten, M., Bos, A., Kidron, M., Kamphues, F., Kuiper, S., Jonker, W., Ryan, A., and Hinz, P., ``Progress report on the integration of the irtf adaptive secondary mirror,'' in [{\em Adaptive Optics for Extremely Large Telescopes 8 (AO4ELT8)}{\nolinebreak\hspace{0.1em}]},  (2025).

\bibitem{dinh2024measuring}
Dinh, C.~K., Rayner, J.~T., Lockhart, C., Chun, M., and Connelley, M., ``Measuring irtf image quality and modeled improvement through use of felix and the prototype adaptive secondary mirror,'' in [{\em Ground-based and Airborne Telescopes X}{\nolinebreak\hspace{0.1em}]},   {\bf 13094},  1763--1779, SPIE (2024).

\bibitem{woillez2019naomi}
Woillez, J., Abad, J., Abuter, R., Carpentier, E.~A., Alonso, J., Andolfato, L., Barriga, P., Berger, J.-P., Beuzit, J.-L., Bonnet, H., et~al., ``Naomi: the adaptive optics system of the auxiliary telescopes of the vlti,'' {\em Astronomy \& Astrophysics}~{\bf 629},  A41 (2019).

\bibitem{lai2021crime}
Lai, O., Chun, M., Dungee, R., Lu, J., and Carbillet, M., ``Do-crime: dynamic on-sky covariance random interaction matrix evaluation, a novel method for calibrating adaptive optics systems,'' {\em Monthly Notices of the Royal Astronomical Society}~{\bf 501}(3),  3443--3456 (2021).

\bibitem{rayner2003spex}
Rayner, J.~T., Toomey, D., Onaka, P., Denault, A., Stahlberger, W., Vacca, W., Cushing, M., and Wang, S., ``Spex: a medium-resolution 0.8--5.5 micron spectrograph and imager for the nasa infrared telescope facility,'' {\em Publications of the Astronomical Society of the Pacific}~{\bf 115}(805),  362 (2003).

\bibitem{1983PASP...95..163K}
{King}, I.~R., ``{Accuracy of measurement of star images on a pixel array.},'' {\em Publications of the Astronomical Society of the Pacific}~{\bf 95},  163--168 (Feb. 1983).

\end{thebibliography}
\bibliographystyle{spiebib} 

\end{document}